\documentclass{elsart3}
\usepackage{natbib}
\usepackage{amssymb}
\usepackage{graphicx}
\usepackage{hyperref}

\def\asec{\ifmmode ^{\prime\prime}\else$^{\prime\prime}$\fi}

\def\degs{\ifmmode ^{\circ}\else$^{\circ}$\fi}
\def\degr{\ifmmode ^{\circ}\else$^{\circ}$\fi}
\def\amin{\ifmmode ^{\prime}\else$^{\prime}$\fi}
\def\arcsec{\ifmmode ^{\prime\prime}\else$^{\prime\prime}$\fi}
\def\farcs{\hbox{$.\!\!^{\prime\prime}$}}  

\def\degs{\ifmmode ^{\circ}\else$^{\circ}$\fi}
\def\amin{\ifmmode ^{\prime}\else$^{\prime}$\fi}
\def\farcm{\hbox{$.\mkern-4mu^\prime$}}

\begin{document}


\unitlength=1mm
\def\EE#1{\times 10^{#1}}
\def\gcm{\rm ~g~cm^{-3}}
\def\cm3{\rm ~cm^{-3}}
\def\kms{\rm ~km~s^{-1}}
\def\cms{\rm ~cm~s^{-1}}
\def\ergs{\rm ~erg~s^{-1}}
\def\wl{~\lambda}
\def\wll{~\lambda\lambda}
\def\Nii{M(^{56}{\rm Ni})}
\def\FeI{{\rm Fe\,I}}
\def\FeII{{\rm Fe\,II}}
\def\FeIII{{\rm Fe\,III}}
\def\Niii{M(^{57}{\rm Ni})}
\def\FeIb{{\rm [Fe\,I]}}
\def\FeIIb{{\rm [Fe\,II]}}
\def\FeIIIb{{\rm [Fe\,III]}}
\def\OIb{{\rm [O\,I]}}
\def\OIIb{{\rm [O\,II]}}
\def\OIIIb{{\rm [O\,III]}}
\def\SIIb{{\rm [S\,II]}}
\def\ArIIIb{{\rm [Ar\,III]}}
\def\NiIIb{{\rm [Ni\,II]}}
\def\Msun{~{\rm M}_\odot}
\def\Ti44{M(^{44}{\rm Ti})}
\def\MZA{M_{\rm ZAMS}}
\def\mum{\mu{\rm m}}
\def\muJ{\mu{\rm Jy}}
\def\psr{PSR~B0540-69.3}
\def\snr{SNR~0540-69.3}

\def\lsim{\!\!\!\phantom{\le}\smash{\buildrel{}\over{\lower2.5dd\hbox{$\buildrel{\lower2dd\hbox{$\displaystyle<$}}\over\sim$}}}\,\,}
\def\gsim{\!\!\!\phantom{\ge}\smash{\buildrel{}\over{\lower2.5dd\hbox{$\buildrel{\lower2dd\hbox{$\displaystyle>$}}\over\sim$}}}\,\,}

\begin{frontmatter}

\title{Optical observations of the young supernova remnant SNR~0540-69.3 and its pulsar\thanksref{aaa}}

\thanks[aaa]{Based on
observations performed at the European Southern Observatory, La Silla and
Paranal, Chile (ESO Programmes 56.C-0731 and 67.D-0519).}


\author[1,2]{N.~I.~Serafimovich}, \author[2]{P.~Lundqvist}, 
\author[1]{Yu.~A.~Shibanov}, \author[2]{J.~Sollerman} 

\address[1]{Ioffe Physical Technical Institute, Politekhnicheskaya 26, 
St. Petersburg, 194021, Russia}
\address[2]{Stockholm Observatory, AlbaNova Science Center, Department 
of Astronomy, SE-106 91 Stockholm, Sweden}

\begin{abstract}
We have used the ESO NTT/EMMI and VLT/FORS1 instruments to examine the 
LMC supernova remnant 0540-69.3 as well as its pulsar (PSR B0540-69) and 
pulsar-powered nebula in the optical range.
 Spectroscopic observations of the remnant covering the range 
of $3600-7350$~\AA\ centered on the pulsar produced results consistent 
with those of Kirshner et al. (1989), but also revealed many new emission
lines. The most important are [Ne~III]~$\lambda\lambda$3869, 3967 and Balmer
lines of hydrogen. In both the central part of the remnant, as well as in 
nearby H~II regions, the [O~III] temperature is higher 
than $\sim 2\times10^4$~K, but lower than previously estimated. For 
PSR B0540-69, previous optical data are mutually inconsistent: 
HST/FOS spectra indicate a significantly higher absolute flux and steeper 
spectral index than suggested by early time-resolved groundbased UBVRI 
photometry. We show that the HST and VLT 
spectroscopic data for the pulsar have $\gsim 50$\% nebular contamination, 
and that this is the reason for the previous difference. Using HST/WFPC2 
archival 
images obtained in various bands from the red part of the optical  
to the NUV range we have performed an accurate photometric study
of the pulsar, and find that the spectral energy distribution of the 
pulsar emission
has a negative slope with $\alpha_{\nu} = 1.07^{+0.20}_{-0.19}$. 
This is steeper than derived from previous
UBVRI photometry, and also different from the almost flat spectrum of the
Crab pulsar.    
We also estimate that the proper 
motion of the pulsar is $4.9\pm2.3$ mas yr$^{-1}$, corresponding to a 
transverse velocity of $1190\pm560 \kms$, projected along the southern 
jet of the pulsar nebula.
\end{abstract}

\begin{keyword}
supernova remnants -- pulsars -- spectroscopy -- photometry -- astrometry --
supernova remnants: individual: SNR 0540-69.3 -- pulsars: individual: PSR B0540-69.3  


\end{keyword}
\end{frontmatter}

\section{Introduction}

Supernova remnant (SNR) 0540-69.3 (henceforth simply 0540) has been 
observed at wavelengths ranging from X-rays to the radio. Both in the radio 
\cite{Man93b} 
and in X-rays  \cite{SH94,GW00}, 
the remnant  is bounded by an outer shell, which has a radius of at 
 least $\sim 30\arcsec$. 
\begin{figure*}[htb]
\begin{center}
\includegraphics[width=110mm, clip]{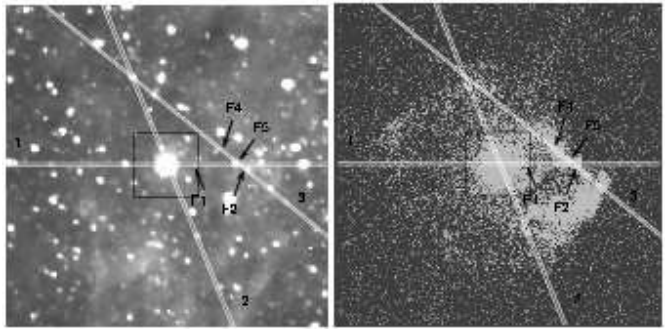}
\end{center}  
\caption{A 100\arcsec$\times$100\arcsec\ field around 0540. 
The left panel is an optical image obtained with the ESO NTT 
in a narrow band filter centered on the [O~III]~($\lambda=5007$~\AA)  
emission line,   
and the right panel shows an X-ray image obtained with Chandra 
in the 1.5$-$6.4 keV range.
White parallel lines marked by numbers show the slit positions used in  
our spectroscopic observations of the SNR with 
the NTT 
(slit 2) and VLT 
(slits 1 and 3). 
The areas marked ``F1, F2, F4, F5" show the positions of several 
emission-line regions identified from the spectroscopic
observations and described in detail in \cite{Ser05}.
A 20\arcsec$\times$20\arcsec\ box marks the inner region of 
the remnant enlarged 
in Fig.~2.  
}\label{f:OIIIim}
\end{figure*}
Inside the outer shell, the emission from the remnant is concentrated to a
substantially smaller nebula  \cite{Mat80,Car92}. In [O~III]~ 
($\lambda=5007$~\AA), the diameter is $\sim 8\arcsec$ (henceforth we will 
refer to this as the `central part of the SNR' (SNRC)). In this and other 
lines, the emission appears to mainly come from a few blobs and filaments. 
The SNRC also emits an optical continuum, believed
to be synchrotron emission from the pulsar wind nebula (PWN) \cite{Cha84, %
Ser04}. Such a nebula is expected since the remnant 
harbors the young pulsar PSR B0540-69 which is observed to emit in 
X-rays \cite{SHH84}, in the optical \cite{MP85}, and in the 
radio \cite{Man93a}. The SNRC with its pulsar bears many 
similarities to the Crab Nebula, which is why 0540 is sometimes 
referred to as the Crab's twin. A detailed comparison of PSR B0540-69 with 
the Crab pulsar and its PWN can be found in \cite{Ser04}. 

The spectrum of the SNRC is dominated by forbidden oxygen and 
sulphur lines. From the strong oxygen lines it has been classified as 
an ``oxygen-rich SNR'' (OSNR). This class of objects has a handful of members,
of which Cas A is the most studied object 
(e.g.,~\cite{Morse04} and references therein). However, Cas A
is not pulsar-powered, although a point-like X-ray source has 
been detected near its center \cite{tan99}. 
Thus 0540 is an interesting link between pulsar-powered remnants
and OSNRs. 
Here we present some new results for this object obtained with the
ESO NTT and VLT, and with the HST. 
%
\section{Observations}
%
Observations of 0540 were performed on 1996~January 17, using 
the ESO NTT telescope with the ESO Multi-Mode Instrument 
(EMMI)\footnotemark \footnotetext{http://www.ls.eso.org/lasilla/Telescopes/NEWNT}.  
Narrow-band images in [O~III]~($\lambda=5007$~\AA) were obtained 
using a zero velocity  
[O III]$/$0 filter\footnotemark \footnotetext{http://www.ls.eso.org/lasilla/Telescopes/NEWNT/
emmi/emmiFilters.html}   
and an exposure time of 30 minutes.
The pixel size was 0\farcs268. 
We also carried out low-resolution (2.3 \AA$/$pixel)
long-slit spectroscopy of 0540 in the 3850-8450~\AA\ range.
The [O~III]$/$0 image 
is shown in Fig.~\ref{f:OIIIim} ({\it left}), where we also show 
the slit position of the NTT spectral observation (marked by ``2'').
%
%
Further spectroscopic observations were carried  out in the 
3600--6060~\AA\ range on 2002 January 9 and 10 
with the ESO VLT telescope using  the 
FOcal Reducer/low dispersion Spectrograph 1
(FORS1)\footnotemark \footnotetext{http://www.eso.org/instruments/fors1/} with   
a dispersion of 1.18~\AA$/$pixel, and a spatial scale 
of 0\farcs2 per pixel. 
Total exposure times of 154 and 132 minutes were used for 
the two slit positions, 
marked by ``1'' and  ``3'' in Fig.~\ref{f:OIIIim}, respectively. 
Slits 1 and 2 were chosen to include the SNRC and the pulsar. 
Slit 3 does not cross the SNRC, but was placed
to probe emission from the outer shell in a region
where it is most clearly identified in the Chandra X-ray image \cite{Hw01}
(Fig.~\ref{f:OIIIim}, right). 
The Chandra image in Fig.~\ref{f:OIIIim} was obtained on 1999 November 
22 with the ACIS-S in the $1.5-6.4$ keV range with a spatial resolution 
of 0\farcs492 per pixel using a total exposure time of 27.8 ks.  
All spectroscopic observations were performed using a 
slit width of~1\arcsec~and the seeing was generally about 1\arcsec. 
%

\begin{figure*}[htb]
\begin{center}
\includegraphics[width=80mm, clip]{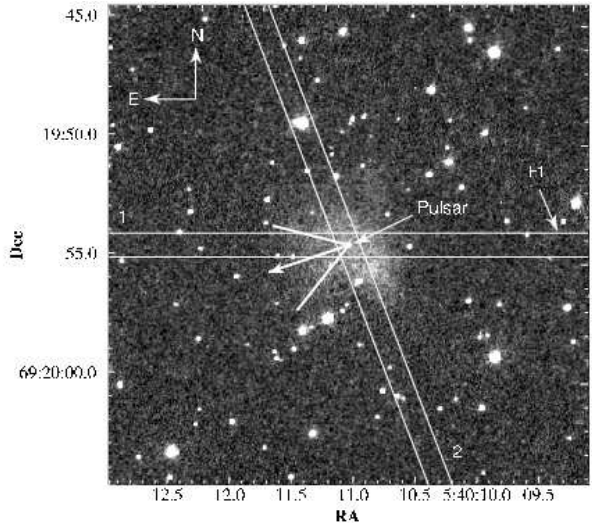}
\end{center}  
\caption{A 20\arcsec$\times$20\arcsec\ HST/WFPC2 image of the 
inner region of 0540 in the F547M band.
Slit positions 1 and 2 shown in Fig.~\ref{f:OIIIim}, as well as the position of 
the pulsar and the emission-line filament F1 are marked. The slit widths 
are 1\arcsec. The diffuse emission surrounding the pulsar comes from the PWN. 
Note its elongation in the NE-SW direction. 
The proper motion of the pulsar is in the south-east direction, and 
its $1\sigma$ $-$ angular uncertainties are shown by the thick arrow and 
lines (see text). Coordinates refer to epoch J2000. 
}\label{f:F547im}
\end{figure*}

The 0540 field has also been imaged with the HST/WFPC2 on 
several occasions in various bands. Data for observations through the 
F336W, F502N, F547M, F673N and F791W filters were obtained on 
1999 October 17 using total exposure times of 600~s, 11000~s, 800~s, 
8200~s and 400~s, respectively  \cite{Mo03}. 
Data in the narrow band F658N and wide band F555W filters were obtained on 
1995 October 19 with 4000~s, and 600~s exposures, respectively \cite{Car00}. 
We retrieved these images from the HST archive and used them in our analysis.
The SNRC was exposed on the Planetary Camera (PC) chip with 
a spatial resolution of 0\farcs046 per pixel revealing 
the pulsar and the structure of its PWN (Fig.~\ref{f:F547im}),
as well as the fine structure of the SNRC 
and its neighborhood. All the data obtained have been reduced and 
calibrated in a standard way (for details see  \cite {Ser04, Ser05}). 

\section{Results}
%
%
\subsection{Spectroscopy of the central part of 0540.}  
\begin{figure*}[htb]
\begin{center}
\includegraphics[width=120mm, clip]{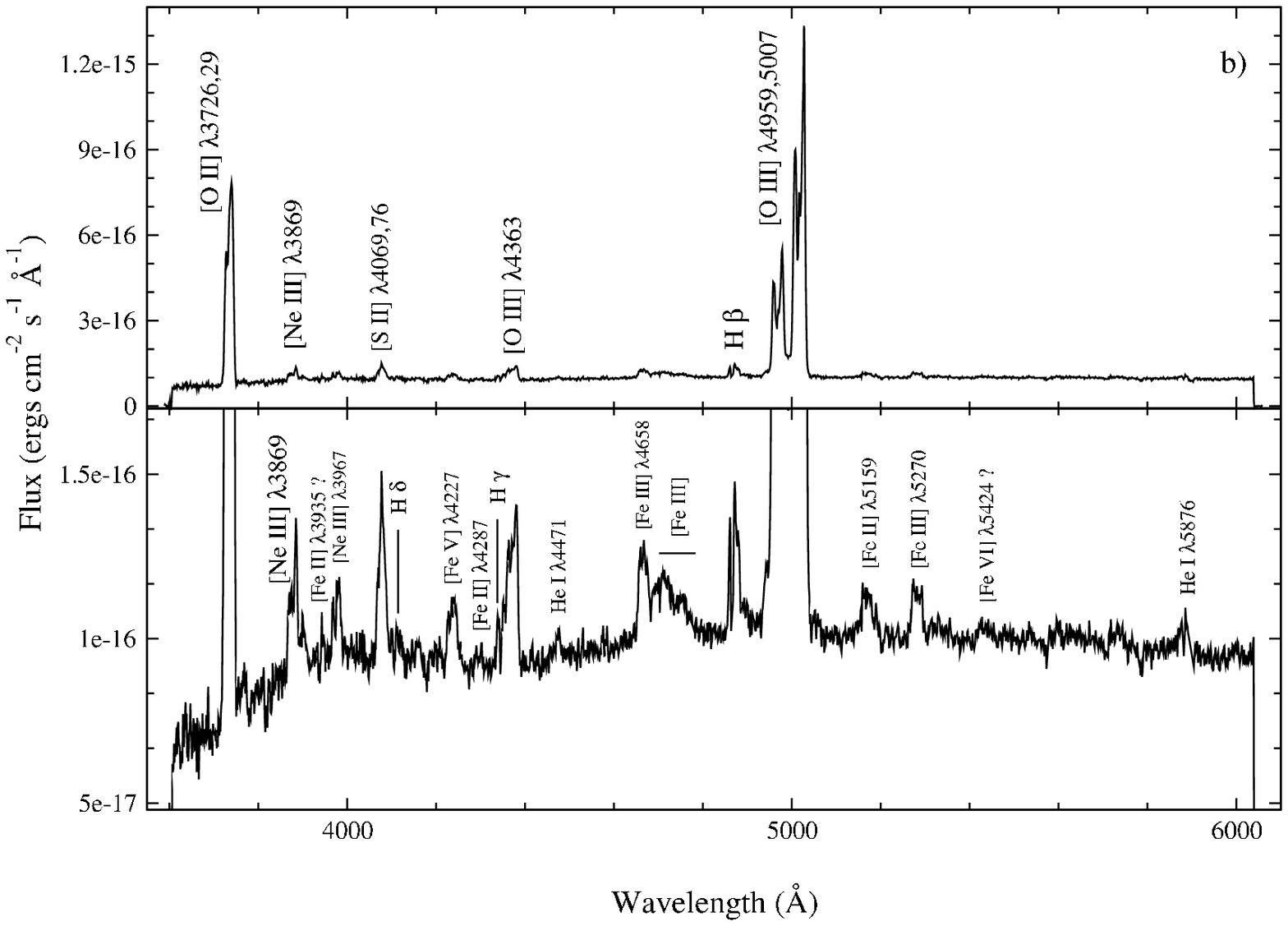}
\end{center}  
\caption{{\it Top}: The ESO VLT slit 1 spectrum of the central part of 
0540 (marked ``b''). {\it Bottom}: Same spectrum but with an expanded flux 
scale to highlight weaker lines. The spectrum has been dereddened using 
$E(B-V)=0.20$ and R=3.1. The high signal-to-noise ratio of the VLT 
spectrum made it possible to detect faint lines from the SNRC.
}\label{f:Snrs_sp}
\end{figure*}

Spectroscopic studies of the SNRC were carried out in detail 
by Kirshner et al. (1989) \cite{Kir89} using a larger slit width ($1\farcs5$) 
and at a different position angle. To compare our results with theirs,
we extracted 1D spectra of the SNRC from our spectral images using
the {\sf IRAF}
procedure {\sf apall} and spatial extents of 10\arcsec\ and 8\arcsec\ centered
on the pulsar for slits ``1" and ``2", respectively. The extracted windows
correspond to the observed extents of the SNRC along the respective slit
directions. The extracted VLT spectrum from slit 1 shown in 
Fig.~\ref{f:Snrs_sp} is consistent with that of Kirshner et al.~\cite{Kir89}.
However, the higher sensitivity and spectral resolution of the 
VLT observations also allow us to find many new lines not previously
detected. The most important findings are [Ne~III]$\wll$3869, 3967 and 
Balmer lines of hydrogen all the way down to H~I~$\lambda$3889.  
These lines are marked in Fig.~3 and detected at least at $\ge 5\sigma$ 
significance level. They belong to the SNR material, and not to the LMC 
background, since the measured shifts of their line centroids correspond to 
a velocity of~$500-800\kms$, which is consistent with the velocities of the 
much stronger [O~III] and [S~II] lines. Moreover, the lines are velocity 
broadened to the same extent (a typical width is $\sim 2000 \kms$) as other 
lines emitted by the remnant. While the neon lines can be used to derive an
O/Ne ratio which in turn constrains the progenitor mass (cf. \cite{Ser05}), 
the Balmer lines show that the previously detected 
emission around H$\alpha$ is at least partly due to H$\alpha$, and not only 
to [N~II]~$\lambda$6583 or any other line as previously 
discussed \cite{Car00, Kir89}.


\subsection{SNRC density from [S~II]~$\lambda\lambda$6716,~6731.}
The intensity ratio $R_{\rm [S~II]} = \frac{I(\lambda6716)}{I(\lambda6731)}$  
is sensitive to the electron number density, $N_{\rm e}$ 
(e.g., \cite{Oster89}). 
As discussed in \cite {Ser05}, the two components of [S~II] detected
using slit 2, blend together 
because of the velocity broadening of the emitting gas in 0540. 
There are, however, a few positions along the slit 
for which a deblending is possible. 
From various fits to the line profiles at these positions, 
we obtained $R_{\rm [S~II]} \approx 0.7\pm0.1$. 
To estimate the density, we used the multilevel model for S~II described 
in \cite{LF96}, but with atomic data further updated 
according to \cite{RBS96}.  
Adopting the above $R_{\rm [S~II]}$ value 
we obtain $N_{\rm e} = (1.4-4.3)\EE3 \cm3$ assuming the temperature $T=10^4$ K 
and $N_{\rm e} = (1.8-5.3)\EE3 \cm3$ at $T=2\EE4$ K. The lower temperature
is probably more likely considering our findings for [O~III] (see below),
so a reasonable density range 
is $N_{\rm e} = (1-5)\EE3 \cm3$. This is similar to the density in the
Crab Nebula for which [O~II] and [S~II] line ratios indicate electron
densities in the range of $4\EE2 - 4\EE3 \cm3$ for various
filaments observed  \cite{DF85}.

\subsection{SNRC temperature from [O~III].} 
As can be seen from Fig. 3, [O~III]$\wl$4363 is clearly detected in the
spectrum of the SNRC of 0540. This enables us to estimate the temperature 
from the flux 
ratio $R_{\rm [O~III]} = \frac{I(\lambda\lambda4959,5007)}{I(\lambda4363)}$,
(e.g., \cite {Oster89}). A complication for 0540 is, however, that lines from 
the SNRC are blended due to velocity broadening. In particular, 
[Ne~III]$\wll$3869, 3967 and [O~III]$\wl$4363 are contaminated by H~I 
lines. This leads to an underestimate of $R_{\rm [O~III]}$, and consequently
an overestimate of the temperature if deblending is not made. 
To deblend these lines we used H$\beta$ as a 
template profile for all H~I lines. Incorporating a six-level model atom 
for O~III from \cite{Mar00} gives $T_{\rm e} \approx 2.4\EE4$~K (for the 
electron density we found above from [S~II]). Although our [O~III] temperature 
is significantly lower than that of \cite{Kir89}, who 
estimated $T_{\rm e} \sim 3.4\EE4$~K without taking deblending into account, it
is still much higher than in normal H~II regions. The corresponding
temperature in the Crab Nebula is in the range $(1.1-1.8)\EE4$~K \cite{DF85}, 
and is thought to mainly arise from photoionization heating \cite{Kir89}. 
Our value of $T_{\rm e}$ from [O~III] may point in the
direction of shock heating not being the sole source of heating of the SNRC of
0540 since photoionization by a hard spectrum, like in the Crab, can also 
give rise to high [O~III] temperatures (e.g., \cite{Kir89}). Modeling is
needed to sort out which source of heating dominates in the SNRC of 0540.

\subsection{0540 filaments.} 
%
0540 is near the LMC H~II region DEM 269 and the 
OB association LH 104 \cite{Kir89}. 
This is reflected in our spectroscopic images which contain strong, 
narrow emission lines that vary spatially in strength along the slits. 
These lines are significantly narrower (actually unresolved) than the lines 
from the SNRC of 0540.
To study whether or not some of the H~II (or at least emission-line) 
regions are affected by the SNR activity, we selected several filaments 
detected along our slits and analyzed their spectra. 
Many of the filaments actually do show highly ionized
ions (as mainly deduced from iron) as expected for regions close to the
X-ray emitting parts of the remnant.
The filaments we studied in greater detail were along the VLT slits ``1" 
and ``3", and they are marked F1, F2, F4, F5  in Fig.~\ref{f:OIIIim}. 
For reference we also looked at a filament (named F3) which in 
projection is situated 1\farcm4 from the pulsar (to 
the east, and outside the region shown in Fig.~1). While F3 can still be 
affected by the radiation field of 0540, it clearly cannot be shock excited 
by the remnant. 
%

Filament F1, which in projection is the closest one to the SNRC, is 
remarkably hot, $T_{\rm e} \sim 3.7\EE4$~K. Filaments F2, F4 and F5 are 
closer to the outer shock front detected in X-rays (as seen in 
projection), and they all have temperatures in excess of $2.3\EE4$ K. The 
reference filament F3 has an [O~III] temperature of $\sim 1.7\EE4$ K. 
The temperatures of all filaments (even that of F3) are substantially 
higher than in normal H~II regions. Despite the fact that LMC is less 
metal-rich than the Milky Way, and metal line cooling in H~II regions 
is therefore expected to be less efficient there, typical [O~III] 
temperatures in, e.g., the 30 Doradus region vary rather mildly ($\pm 140$~K)
from the average value of 10\,270~K \cite{KC02}. The high 
temperature of filaments F1-F5 around the SNRC of 0540 suggests that they 
are all affected by the remnant. 

The spectral resolution of our FORS1 observations is $\sim 180 \kms$ at 
[O~III]~$\lambda$5007~\AA, but the line center of the filament lines can be
determined to within a fraction of this ($\sim 20 \kms$) for the strongest 
lines. Taking filament F3 as a reference, all the other filaments around 0540
are consistent with having their line center velocities within 
this $\pm 20 \kms$ uncertainty range. In the young remnant of SN~1987A, the 
shocked gas in the inner ring has displacements that are marginally larger
than $20 \kms$ \cite{Pun02}, but the effect is certainly not large enough 
for us to rule out shocks as the excitation mechanism for the 0540 filaments. 
SN~1987A also shows that the draping of the blast wave around the ring 
clumps produces line profiles with widths that are in excess of the spectral 
resolution of our 0540 observations. This may pose a stronger constraint on 
the shock hypothesis for the 0540 filaments than the lack of line displacement, 
as we do not see any broadening (beyond the instrumental broadening) of the 
0540 filament lines \cite{Ser05}. We note, however, that both the line 
displacement and the line broadening are geometry dependent, and the geometry 
may certainly be less extreme in the 0540 filaments than in the ring clumps 
of SN~1987A. A shock imprint on the emission lines may therefore be less 
obvious for 0540 than for SN 1987A. An alternative, or 
maybe complementary, excitation scenario for filaments F1, F2, F4 and F5
around 0540 is that they are photoionized by a time-varying EUV/X-ray source. 
It was shown in \cite{LF96} that [O~III] temperatures of several times $10^4$~K 
can be attained in such a case. The ionizing source could in this case be 
naturally provided by the X-rays created when the blast wave overtakes 0540 
filaments. Modeling to test the photoionization scenario is in 
progress \cite{Ser05}. Observations with better spectral resolution will 
also help to resolve the excitation mechanism.
%
%

\subsection{The spectral energy distribution of the pulsar emission.}
\begin{figure*}[t]
\begin{center}
\includegraphics[width=100mm, height=126mm, clip, angle=0]{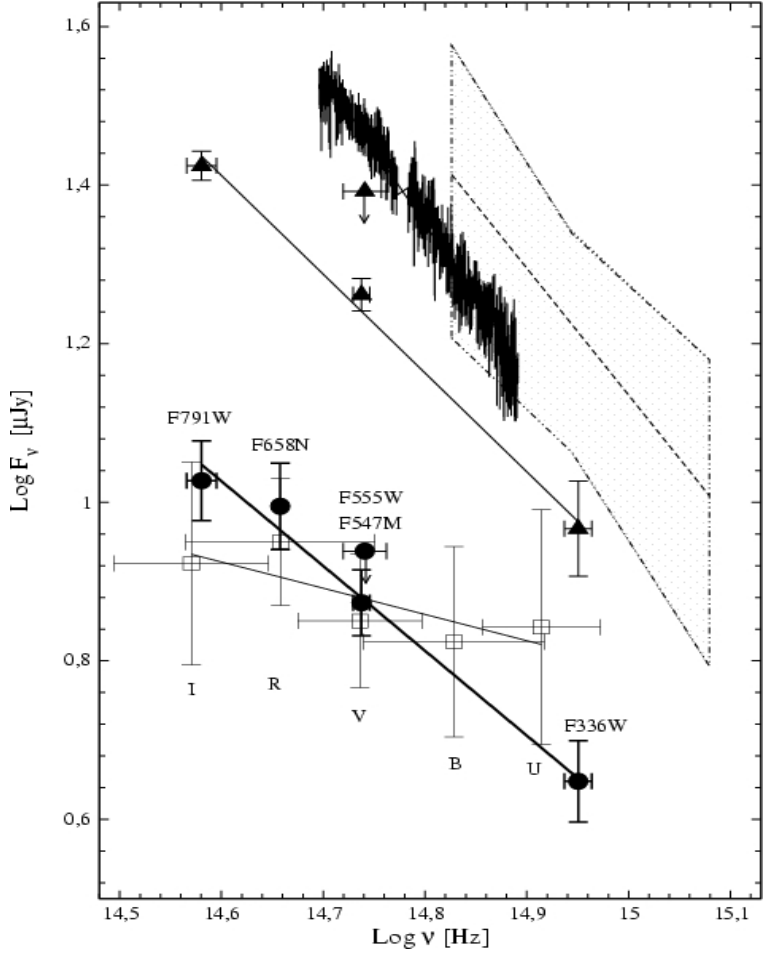}
\end{center}
\caption{The optical spectrum of \psr\ obtained with different telescopes and 
instruments. The uppermost is the VLT slit 1 spectrum for the 6-pixel area 
centered at the pulsar. 
The bright [O~III] nebular lines have been removed.  
The dashed line and the hourglass-shaped error-box show a power law fit 
and $1\sigma$-uncertainties of the UV spectrum obtained by Hill et al. (1997).  
Filled triangles show the HST photometry with a 10-pixel circular
aperture to compare with the above spectra.
Filled ellipses show our HST photometric results, see \cite{Ser04}. 
Open rectangles are the photometric UBVRI data from Middleditch et al.~(1987).
All data are dereddened using $E(B-V)=0.20$. Solid lines show power-law 
fits to the HST and Middleditch photometric data sets. 
}\label{f:PSRopt}
\end{figure*}

\psr~is one of a few pulsars for which a near-UV and optical spectrum has been
reported. Hill et al. \cite{Hill97} obtained a time-integrated near-UV 
spectrum with HST/FOS and Middleditch et al. \cite{Midd87}
used time-resolved photometry to establish a broadband ground-based UBVRI 
`spectrum' in the optical. 
These two investigations show, however, a significant difference in the 
absolute flux in the spectral range where they overlap. To check this mismatch 
we used two recent sets of data, one is our VLT/FORS spectroscopy 
of 0540, and the other is the HST/WFPC2 imaging \cite {Car00, Mo03}. 
Our spectroscopic and photometric results are shown in 
Fig.~\ref{f:PSRopt} \cite {Ser04}.  
Considered separately, they are in good agreement with previous results. 
However, they  do not erase the significant discrepancy between 
the spectroscopic and photometric data sets, which is much 
larger than the statistical uncertainties of our measurements. 
The only plausible explanation to this discrepancy
is that the HST/FOS spectroscopy is strongly contaminated by the PWN.
As an additional test we measured the pulsar flux in the F336W, F547M, F555W,
and F791W bands, using a circular aperture with a radius of 10 PC-pixels 
(total diameter of 0\farcs92) centered on the pulsar, without subtraction 
of the background. 
These conditions reproduce the parameters of the spectral measurements 
within a circular aperture of almost the same diameter (0\farcs86) 
made with the HST/FOS by Hill et al.~1997 \cite{Hill97}. The 10-pixel 
fluxes are much closer to the HST and VLT spectral fluxes. We estimate that 
within a 10 pixel radius, the PWN contributes at least 50\%. This can also 
be seen by comparing our accurate pulsar photometry with our 10 pixel 
test \cite{Ser04}. We conclude that the previously published spectral data 
on the pulsar emission cannot be considered as reliable. 
Our broadband HST `spectrum' (or rather broad-band spectral energy 
distribution) of the pulsar, where the background from the nebula has been 
accurately subtracted off, can be considered as a fair estimate of the pulsar 
spectral energy distribution. It has a nonthermal origin and is described 
by a power law, $F_{\nu} \propto \nu^{-\alpha}$, with the spectral 
index $\alpha_{\nu} =1.07_{-0.19}^{+0.20}$, while that of Middleditch et al. 
has $\alpha_{\nu} =0.33\pm0.45$ using updated dereddening corrections. 
The flatter spectrum of Middleditch et al. \cite{Midd87} is partially due 
to a spectral jump upwards for the U band that could be due to a systematic 
flux error in this band.
%
%
%

\subsection{Proper motion of \psr.}
The position of \psr\ is defined on the HST PC chip frames with an accuracy
of better than 0.17 PC pixels which corresponds to 0\farcs0077. 
This allows a direct estimate of the proper motion of the pulsar using an 
accurate superposition of the F555W and F547M images taken at 
epochs separated by 3.995 years. Using the positions of 9 reference 
stars to construct the coordinate transformation between the two images, 
we find a proper motion $\mu=4.9\pm2.3$ mas yr$^{-1}$  
in the South-East direction (see Fig.~\ref{f:F547im}) at a position angle 
of $109^{\rm o}\pm33^{\rm o}$ (projected along the southern jet of its PWN). 
The significance of this result is low and can be considered  
only as an attempt to make a first direct measurement of the proper motion. 
Based on the displacement between the pulsar optical position and 
the center of the PWN, as seen in radio,  
Manchester et al.~1993b \cite{Man93b} argued for a similar value of the proper 
motion but in the South-West direction  (in the plane of the PWN torus, as 
seen in projection). We note that the proper motion of the Crab 
pulsar is projected along the symmetry axis of the inner Crab nebula, as 
defined by the direction of the PWN jet discovered by ROSAT,  
and that a similar situation applies to the Vela pulsar \cite{Luca00}. 
If our estimates are confirmed, we have the intriguing situation 
that all these three young pulsars move along their jet axes. 
However, while the Crab and Vela pulsars both have 
transverse velocities of $\sim 130\kms$, our results for \psr\ 
indicate a much higher value of 1190$\pm560\kms$, assuming
a distance to the LMC of 51 kpc. A third epoch of HST imaging is 
clearly needed to establish this result at a higher significance level.\\

This work has been partially supported by Russian Foundation of Basic 
Research (RFBR,   
grants 02-02-17668, 03-02-17423 and 03-07-90200), Russian Leading Scientific School (RLSS) 
program 1115.2003.2, and by The Royal Swedish Academy of Sciences.
The research of PL is sponsored by the Swedish Research Council.
PL is a Research Fellow at the Royal Swedish Academy supported by a grant 
from the Wallenberg Foundation. We also acknowledge support from the Swedish
Research Council.

{}
\end{document}